# Approaches to developing tolerance and error budget for active three mirror anastigmat space telescopes

Heejoo Choi[a,b], Young-Sik Kim[a], Hyukmo Kang[a], Solvay Blomquist[a], Hill Tailor[a], Douglas Kelly[c], Mike Eiklenborg[c], Ewan S. Douglas[c], Daewook Kim[a,b,c,*]

[a]Wyant College of Optical Sciences, University of Arizona, 1630 E. University Blvd., Tucson, AZ 85721, USA
[b]Large Binocular Telescope Observatory, University of Arizona 933 N Cherry Avenue, Tucson, AZ 85721, USA
[c]Department of Astronomy and Steward Observatory, University of Arizona, 933 N. Cherry Ave., Tucson, AZ 85721, USA

**ABSTRACT**

The size of the optics used in observatories is often limited by fabrication, metrology, and handling technology, but having a large primary mirror provides significant benefits for scientific research. The evolution of rocket launch options enables heavy payload carrying on orbit and outstretching the telescope's form-factor choices. Moreover, cost per launch is lower than the traditional flight method, which is obviously advantageous for various novel space observatory concepts. The University of Arizona has successfully fabricated many large-scale primary optics for ground-based observatories including the Large Binocular Telescope (LBT, 8.4 meter diameter two primary mirrors), Large Synoptic Survey Telescope (now renamed to Vera C. Rubin Observatory, 8.4 meter diameter monolithic primary and tertiary mirror), and the Giant Magellan Telescope (GMT, 8.4 meter diameter primary mirror seven segments). Launching a monolithic primary mirror into space could bypass many of the difficulties encountered during the assembly and deployment of the segmented primary mirrors. However, it might bring up unprecedented challenges and hurdles, also. We explore and foresee the expected challenges and evaluate them. To estimate the tolerance and optical error budget of a large optical system in space such as three mirror anastigmat telescope, we have developed a methodology that considers various errors from design, fabrication, assembly, and environmental factors.

## 1. INTRODUCTION

The need for larger photon collection areas and higher spatial resolution for space observatory missions is essential for astronomers to survey dimmer and smaller celestial objects. However, the fabrication, installation, and transportation constraints have traditionally limited the size and form factor of space telescopes. Recent advancements in space rocket technologies have started to alleviate these limitations[1], making the realization of a large monolithic primary mirror in space more feasible.

We present a comprehensive evaluation of the tolerance of a large monolithic aperture space telescope. Although our analysis focuses on a generic requirement of space-based observation, the insights gained from this study are relevant and applicable to on-earth telescope systems as well. By conducting a broad initial tolerance analysis, we identified and quantified possible errors, which were then allocated in the optical budget table. The optical budget table plays a pivotal role in guiding each sub-group team within the project. It assists them in meeting stringent requirements while considering the necessary tradeoff to manage the amount of error within acceptable limits. This holistic approach ensures that all components of the telescope work harmoniously to achieve the desired final optical performance.

Moreover, the insights obtained from our tolerance analysis hold good value in the selection of suitable off-the-shelf components that align with the telescope's performance criteria. Making well-considered decisions when choosing these components becomes crucial in upholding the overall cost efficiency during the construction of the telescope.

## 2. TELESCOPE DESIGN OVERVIEW

The Three Mirror Anastigmat (TMA) telescope design is renowned for its benefits in aberration control over a large field of view and adjustability of total size[2]. While reducing the number of mirrors (e.g., Richey-Chretien) may seem like an economical solution for budget constraints, this approach fails to consider the drawbacks it poses to the field of view performance affecting the observation throughput and the complexity of fabrication and assembly.

To address various input factors, including readily available techniques for fabrication, affordable rocket fairing size, and required optical performance, a thorough survey of available design space was conducted. Eventually, the TMA design with a total of 4 mirrors including a folding flat mirror (M4) was chosen. In Figure 1, the placement of M3 almost coincides with the optical surface of M1, ensuring ample space for downstream instruments behind of M1, such as spectrograph and coronagraph. M4 serves as the pupil plane and can be replaced with a deformable and/or fast steering mirror when needed.

The proposed TMA system's configuration, with the intermediate focal/image plane (between M2 and M3) and pupil plane (M4), offers efficient ways to address issues related to stray light. Additionally, this design incorporates an array (e.g., 3 by 5) of off-the-shelf detectors for the telescope (i.e., context camera) focal plane, covering a field of view of 4 × 1.5 mrad. A summary of the optical design values used in this tolerance study is provided in Table 1.

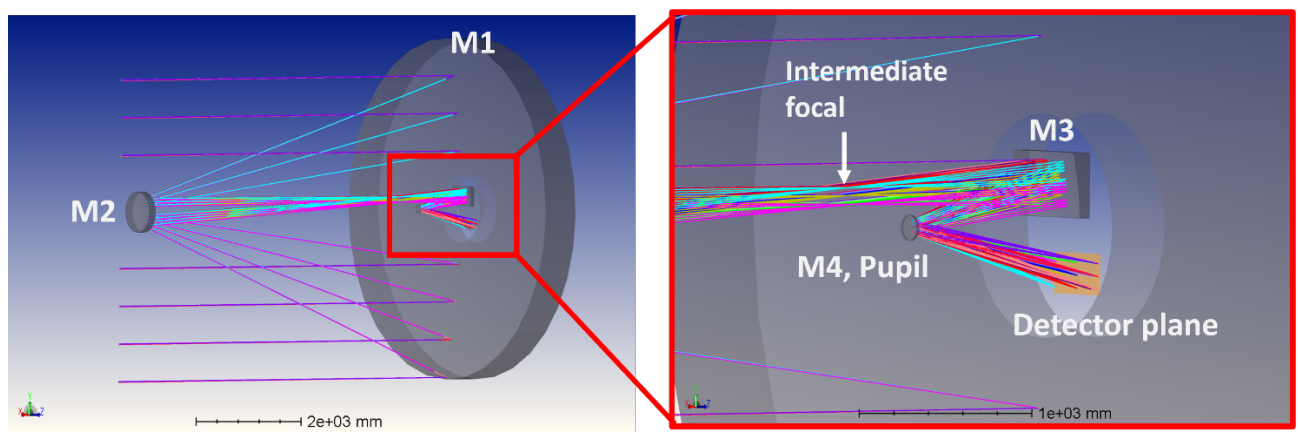

Figure 1. Conceptual TMA telescope optical layout using a 6-meter class primary mirror (M1).

Table 1. The optical design parameter of the benchmark TMS telescope.

| Optical Design Parameter | Nominal Value | Note |
| --- | --- | --- |
| F/# | 15 | |
| Field of View | +/- 0.115×0.043° | Equivalent to 4 × 1.5 mrad full FoV coverage (i.e., 6 mrad$^2$) |
| Field Bias | 0.17° | Center field bias for common optical axis TMA design. |
| Final Focal Plane Size | ~400 × 155 mm | Final Focal Plane Size |
| M1 CA Outer Diameter | 6.42 m | CA is Clear Aperture. |
| M1 CA Inner Diameter | 1.38 m | CA is Clear Aperture. |
| M1 RoC | 16.256 m | RoC is Radius of Curvature. |
| M1 Conic Constant | -0.995357 | |
| M2 Diameter | 780 mm | 760 mm CA diameter. |
| M2 RoC | 1.660575 m | RoC is Radius of Curvature. |
| M2 Conic Constant | -1.566503 | |
| M3 Dimension | 580 × 380 mm | 560 × 360 mm CA. |
| M3 RoC | -1.830517 m | RoC is Radius of Curvature. |
| M3 Conic Constant | -0.718069 | |
| M3 Off-axis Distance | 240 mm | Distance from the parent optical axis to the center of CA. |
| M4 diameter | 150 mm | 130 mm CA diameter. |
| M4 RoC | Inf (i.e., flat) | RoC is Radius of Curvature. |

# 3. TOLERANCE MODEL DEVELOPMENT

During the initial optical design phase, tolerance values are carefully considered to allow for active optics compensation[4]. The nature of the Three Mirror Anastigmat (TMA) design, with more tolerance on M2 position, provides flexibility in accommodating affordable misalignments and aberration compensation. Each mirror's distinct role makes the active optics compensation process easier. Additionally, the tolerance results play a significant role in formulating the fabrication and telescope control plan. The optomechanical design, including mirror supports, the number of hard point supports, active control methods, and thermal control for the observatory, are evaluated in relation to the overall optical performance budget. A comprehensive error analysis is crucial when multiple error sources are expected to exist simultaneously, necessitating an accurate tolerance model emulating mirrors' realistic situation.

To meet the science requirements and observatory environment, the allowed range of misalignment for M1 is established, and a tolerance model is built in Ansys Zemax OpticStudio. For consistency, 500 cases are pre-created using Monte Carlo simulation, and a homemade API algorithm is employed as presented in Figure 2 for further analysis combined with science data processing pipelines. The optical performance for all field points is re-evaluated for each case study, and the results for the 500 cases are shown in full field performance, as depicted in Figure 3.

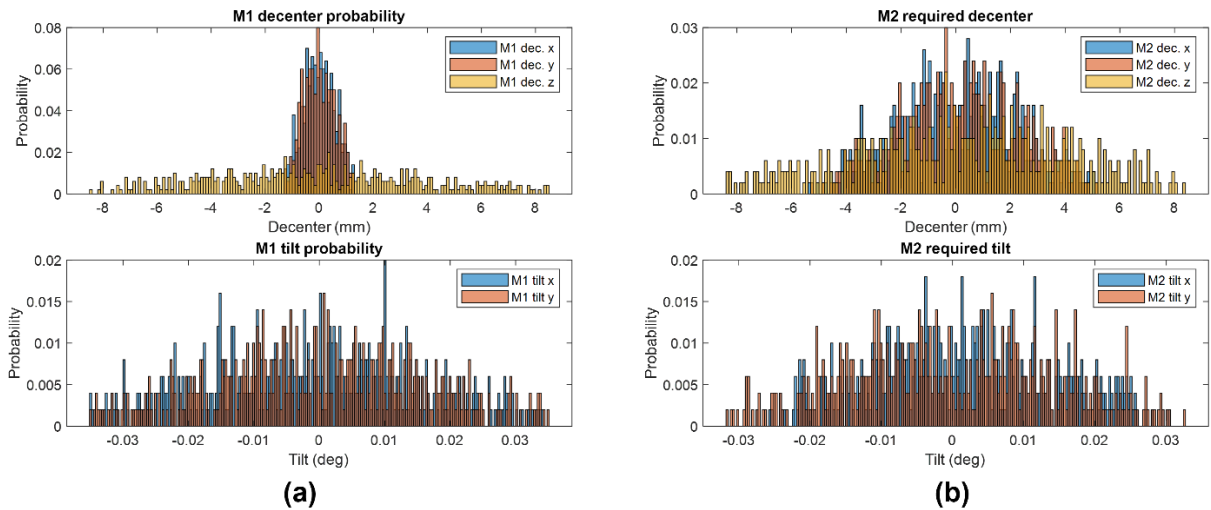

Figure 2. The figure depicts the results of the Monte Carlo simulation, providing two histograms for the case study TMA telescope. In (a), the histogram illustrates the misalignment ranges of M1 for 5 Degrees of Freedom (DoF). The decenter values along the x, y, and z axes are ± 1.3 mm, ± 1.3 mm, and ± 8.5 mm, respectively. The tilt about the x and y axes corresponds to ± 0.035 degrees. In (b), the histogram represents the requirements for M2 active optics motion to maintain diffraction limit performance across the entire field of view. The decenter values along the x, y, and z axes are ± 5.8 mm, ± 5.8 mm, and ± 8.5 mm, respectively. The tilt about the x and y axes corresponds to ± 0.028 degrees and ± 0.035 degrees.

M2 incorporates active optics (e.g., using hexapod) adjustability to control wavefront error resulting from various misalignments. The wavefront error on the final focal plane is assessed using Point Spread Function (PSF) evaluation (e.g., phase retrieval calculation[3]). The optimal position of M2 is determined using Ansys Zemax's optimization tools, ensuring diffraction-limited performance while exploring the maximum affordable misalignment allowance of M1. The findings are presented in Figures 2 and 3.

In wavefront error sensing, limitations in accuracy and time delay between measurement and compensation may exist. Discussions on implementing additional control capabilities (e.g., a deformable M4 and/or M1 hardpoint control akin to LBT primary mirror[5]) are ongoing and aided by the tolerance analysis. A comprehensive tolerance analysis will be conducted once a specific telescope design is chosen and these case study analysis results are finalized. With the current TMA telescope design parameters[6] and the established error range for M1 and the compensation motion for M2 shown in Figure 2, diffraction-limited PSF performance can be achieved throughout the entire field of view, as demonstrated in Figure 3.

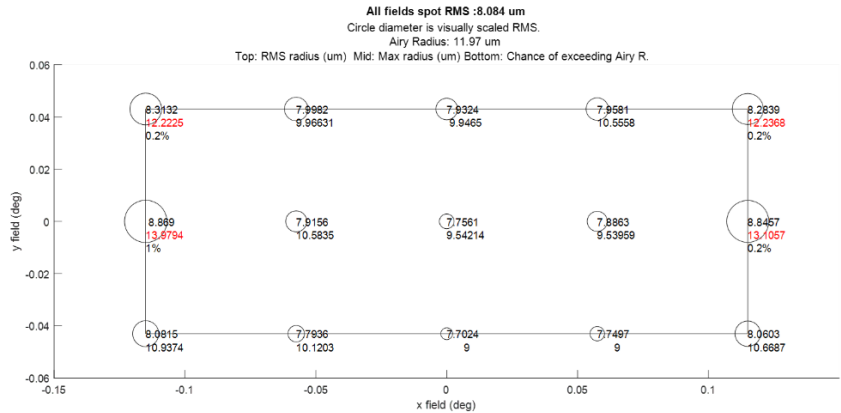

Figure 3. The final performance of the TMA telescope's full field performance. The rectangular box presents the full field of view focal plane and the circle is the scaled PSF size for visualization.

In our case study analysis, we assessed the optical performance at the telescope focal plane across the full field of view, ensuring equal weight on all field points. We acknowledge that this approach might be further relaxed by considering tight or conservative configuration of the 1:2.6 aspect ratio of the field of view, where the horizontal outer edge fields can have a significant impact on tighter tolerance. Nevertheless, our primary objective in setting the baseline tolerance range for each error source is to achieve diffraction limit performance across the full field of view. This allows us to utilize any off-axis PSF at the telescope focal plane for phase retrieval for wavefront sensing and control, reducing the need for unnecessary wavefront error control when placing downstream optics instruments.

The tolerance and strategy of the initial deployment alignment and the daily collimation procedure are currently a subject of ongoing discussion within the observatory's conceptual operations (CONOPS). The specifics of these procedures, including whether they will utilize the same metrology/monitoring system and adjustment knobs, remain under consideration. However, it is anticipated that they will involve distinct loop controls tailored to the engaged downstream optics and instruments.

## 4. OVERALL OPTICAL BUDGET

In the optical error and performance budgeting process, we categorize general error sources into two major categories. The first category involves misalignments of each mirror, which result from relative misalignments and can introduce aberrations. These misalignments stem from factors such as installation misalignments, gravitational relaxations, and thermal drift of the observatory. Interestingly, certain pairs of misaligning mirrors do not negatively impact optical performance due to the balancing of aberrations, and they are considered in our Monte Carlo simulations. The second category comprises irregularities of the optical surfaces, caused by various general sources like fabrication and figuring of optical surfaces, optical testing, hard point support control, gravitational relaxations (gravitational sagging in on-earth observatory case), and thermal gradients over the mirrors. Similar to the first category of errors, the impact of the same amount of irregularity on different mirrors varies depending on their unique aberration balancing roles in the optical design, and these variations are also considered in the Monte Carlo simulation.

Using the comprehensive results from the tolerance analysis, we allocate baseline numbers to each error source while ensuring the total sum remains within predefined case study limitations. This systematic approach allows us to effectively manage and control the various error sources to maintain the desired final optical performance.

To achieve this, we initially categorized the general error sources into five main groups: alignment, optical surface manufacturing, optical surface on orbit, wavefront measurement, and optical design. In the level two subgroup evaluation, we assign certain numbers based on the current best evaluation, which then feeds into the assigned errors to level one. Additionally, user inputs at level three for individual wavefront error sources feed into the current values at level two, which in turn feed into the current values at level one eventually. Any change in the individual error range immediately triggers a recalculation of the comprehensive values, allowing for real-time adjustments and updates to the error budget model. The entire contents of the optical budget table are summarized in Table 2.

Table 2 The general optical budget table assuming the case study TMA telescope design with a three-level evaluation.

| | | | | | | | | |
|---|---|---|---|---|---|---|---|---|
| **STP Error Budget Table (Version - 1, 07/25/2023)** ||||||||||
| **Level-1** ||| **Level-2** ||| **Level-3** |||
| Items | RMS_WFE (nm) | Assigned (%) | Items | RMS_WFE (nm) | Assigned (%) | Items | RMS_WFE (nm) ||
| **I. Alignment** | assigned (RSS) | 13.1 | 9% | **Integration & Testing** | assigned | 11.0 | 52% | M1 misalignment (On-orbit active correction will elliminate this.) | 0.0 |
| | current value (RSS) | 11.6 | | | current value (RSS) | 7.2 | | M3 misalignment (on ground during integration) | 0.0 |
| | margin | 1.5 | | | margin | 3.8 | | M4 misalignment (on ground during integration) | 0.0 |
| | | | | | | | | Context Camera inter-detector array misalignment | 7.2 |
| | | | | **On-orbit Control (Resolution)** | assigned | 5.0 | 24% | M1 hardpoint resolution | 1.4 |
| | | | | | current value (RSS) | 5.4 | | M2 hexapod resolution | 5.2 |
| | | | | | margin | -0.4 | | | |
| | | | | **Thermal & Aging** | assigned | 5.0 | 24% | M1 thermal misalignment (On-orbit active correction will eliminate this.) | 0.0 |
| | | | | | current value (RSS) | 7.2 | | M3 thermal misalignment | 0.0 |
| | | | | | margin | -5.4 | | M4 thermal misalignment | 0.0 |
| | | | | | | | | Context Camera | 7.2 |
| **II. Optical Surface Manufacturing** | assigned (RSS) | 43.2 | 30% | **Manufacturing** | assigned | 43.0 | 91% | M1 figure | 30.0 |
| | current value (RSS) | 42.4 | | | current value (RSS) | 42.4 | | M2 figure | 20.0 |
| | margin | 0.8 | | | margin | 0.6 | | M3 figure | 20.0 |
| | | | | | | | | M4 figure | 10.0 |
| | | | | **Surface Metrology** | assigned | 4.0 | 9% | M1 figure | 0.0 |
| | | | | | current value (RSS) | 0.0 | | M2 figure | 0.0 |
| | | | | | margin | 4.0 | | M3 figure | 0.0 |
| | | | | | | | | M4 figure | 0.0 |
| **III. Optical Surface On-orbit** | assigned (RSS) | 36.3 | 25% | **Opto-mech Support** | assigned | 15.0 | 25% | M1 figure (uncorrectable using M1 bending) | 10.0 |
| | current value (RSS) | 31.0 | | | current value (RSS) | 12.4 | | M2 figure | 5.0 |
| | margin | 5.4 | | | margin | 2.6 | | M3 figure | 5.0 |
| | | | | | | | | M4 figure (M4 is M1 conjugate but uncorrectable using M1 bending) | 2.0 |
| | | | | **Gravity Release** | assigned | 14.0 | 24% | M1 figure (uncorrectable using M1 bending) | 10.0 |
| | | | | | current value (RSS) | 12.4 | | M2 figure | 5.0 |
| | | | | | margin | 1.6 | | M3 figure | 5.0 |
| | | | | | | | | M4 figure (M4 is M1 conjugate but uncorrectable via M1 bending) | 2.0 |
| | | | | **On-orbit Thermal** | assigned | 30.0 | 51% | M1 figure (uncorrectable using M1 bending) | 25.0 |
| | | | | | current value (RSS) | 25.5 | | M2 figure | 3.0 |
| | | | | | margin | 4.5 | | M3 figure | 3.0 |
| | | | | | | | | M4 figure (M4 is M1 conjugate but uncorrectable using M1 bending) | 3.0 |
| **IV. Wavefront Measurement** | assigned (RSS) | 39.0 | 27% | **Measurement Error** | assigned | 20.0 | 32% | Source 1 | 10.0 |
| | current value (RSS) | 152.1 | | | current value (RSS) | 17.3 | | Source 2 | 10.0 |
| | margin | #### | | | margin | 2.7 | | Source 3 | 10.0 |
| | | | | **Wavefront Error Degeneracy** | assigned | 10.0 | 16% | M1 misalignment (on orbit after active correction loop) | 2.9 |
| | | | | | current value (RSS) | 2.9 | | | |
| | | | | | margin | 7.1 | | | |
| | | | | **Time Lag** | assigned | 32.0 | 52% | integration | 150.0 |
| | | | | | current value (RSS) | 151.1 | | computation | 15.0 |
| | | | | | margin | ##### | | actuation | 10.0 |
| **V. Optical Design** | Nominal mean STP RMS WFE (equal FoV weight) | 11.8 | 8% | N/A |||||
| **Net Error** | assigned (RSS) | 70.9 | | Strehl Ratio at 1000 nm wavelength (i.e., 1-sigma^2) |||| 0.80 |
| | current value (RSS) | 161.8 | | Strehl Ratio at 1000 nm wavelength (i.e., 1-sigma^2) |||| -0.03 |
| | margin | -90.9 | | Margin (%) |||| -128.3% |

## 5. CONCLUSIONS

We discussed a comprehensive evaluation of the tolerance of an example large monolithic aperture space telescope, considering case study requirements for a large field of view observation scenarios in space. The insights gained from this case study have broader relevance, extending to on-earth astronomical telescope systems as well. The baseline tolerance analysis outcomes identified and quantified potential errors, effectively allocating them to the optical budget table. This approach ensures that all components of the telescope work together to achieve the desired final telescope performance. Future work will assess the feasibility of these allocations and validate the allocations against an end-to-end system model.

## ACKNOWLEDGMENTS

Portions of this research were supported by the Arizona Board of Regents Technology Research Initiative Fund (TRIF) and by generous donations to the University of Arizona College of Science Steward Observatory.